\documentclass[aps,prd,preprint,superscriptaddress,showpacs,byrevtex]{revtex4}
\usepackage{graphicx}
\usepackage{dcolumn}
\usepackage{bm}
\usepackage{amssymb,amsmath}

\newcommand{\eff}{\varepsilon}
\newcommand{\BR}{{\cal B}}

\newcommand{\pip}{\pi^+}
\newcommand{\pim}{\pi^-}

\newcommand{\etap}{\eta^{\prime}}

\newcommand{\jpsi}{J/\psi}

\newcommand{\EE}{e^+e^-}

\newcommand{\pp}{\pi^+\pi^-}

\newcommand{\beq}{\begin{equation}}
\newcommand{\eeq}{\end{equation}}
\newcommand{\bitm}{\begin{itemize}}
\newcommand{\eitm}{\end{itemize}}

\begin{document}

\title{
\boldmath Measurement of the Matrix Element for the Decay $\eta^{\prime} \to \eta \pi^+\pi^-$}


\author{
{\small
M.~Ablikim$^{1}$, M.~N.~Achasov$^{5}$, L.~An$^{9}$, Q.~An$^{36}$, Z.~H.~An$^{1}$, J.~Z.~Bai$^{1}$, R.~Baldini$^{17}$, Y.~Ban$^{23}$, J.~Becker$^{2}$, N.~Berger$^{1}$
% ...
\\
\vspace{0.2cm}
(BESIII Collaboration)\\
\vspace{0.2cm}
{\it
$^{1}$ Institute of High Energy Physics, Beijing 100049, P. R. China\\
$^{2}$ Bochum Ruhr-University, 44780 Bochum, Germany\\
% ...
$^{5}$ G.I. Budker Institute of Nuclear Physics SB RAS (BINP), Novosibirsk 630090, Russia\\
% ...
$^{9}$ Guangxi University, Naning 530004, P. R. China\\
% ...
$^{17}$ Laboratori Nazionali di Frascati - INFN, 00044 Frascati, Italy\\
% ...
$^{23}$ Peking University, Beijing 100871, P. R. China\\
% ...
$^{36}$ University of Science and Technology of China, Hefei 230026, P. R. China\\
% ...
}} \vspace{0.4cm} }


\begin{abstract}

The Dalitz plot of $\eta^{\prime} \to \eta \pi^+ \pi^-$ decay is studied
using $(225.2\pm2.8)\times 10^6$ $\jpsi$ events collected with the
BESIII detector at the BEPCII $e^+e^-$ collider. With the largest
sample of $\eta^{\prime}$ decays to date, the parameters of the Dalitz plot
are determined in a generalized and a linear representation.  Also
the branching fraction of $\jpsi \to \gamma \eta^{\prime}$ is determined to
be $(4.84\pm0.03\pm0.24)\times 10^{-3}$, where the first error is
statistical and the second systematic.

\end{abstract}


\maketitle

\section{introduction}

Chiral Perturbation Theory is the low energy effective theory of Quantum
Chromodynamics. Below the $\rho$ mass region, the interactions of the
($\pi$, K, $\eta$) particles are systematically analyzed within this framework. The success in the description of these
low-energy interactions makes ChPT a powerful theoretical tool~\cite{pich}.
Although the mass of the $\etap$ is high and $\eta' \to \eta \pp$ decay
has a low Q value, which limit the predictive power of the Effective
Chiral Lagrangian model, the experimental study of the process may
supply information to test the predictions of chiral theory~\cite{ch1, ch2, ch3} and
possible extensions of ChPT such as large-NC ChPT and resonance
Chiral Theory~\cite{RChPT}.
The hadronic decays of the $\etap$ meson
have also been extremely valuable in studies devoted to the effect of the gluon
component~\cite{gluon} and the possible nonet of light
scalars~\cite{nonet}. Previously, the GAMS-4$\pi$ and VES
Collaborations have measured the related Dalitz plot parameters
(GAMS-4$\pi$ for the $\etap \to \eta \pi^0 \pi^0$
channel~\cite{4pi1} and VES for $\etap \to \eta \pp$~\cite{ves1})
complementing older results reported by an early GAMS~\cite{gams}
and CLEO~\cite{cleo} Collaborations. In the isospin limit, the
values of the Dalitz plot parameters should be the same; however the
experimental measurements show some discrepancies among them.

In this article, with a new level of precision, we present results
for the Dalitz plot parameters for $\etap \to \eta \pp$ based on
$(225.2\pm2.8)\times 10^6$ $\jpsi$ events collected by BESIII at
BEPCII.

\section{BESIII and BEPCII}

BESIII/BEPCII~\cite{bepc2} is a major upgrade of the BESII experiment at
the BEPC accelerator~\cite{bepc1} for studies of hadron spectroscopy
and $\tau$-charm physics~\cite{bes3yellow}. The design peak
luminosity of the double-ring $\EE$ collider, BEPCII, is $10^{33}$
cm$^{-2}$s$^{-1}$ at a beam current of 0.93 A. The BESIII detector
with a geometrical acceptance of 93\% of 4$\pi$, consists of the
following main components: 1) a small-celled, helium-based main
draft chamber (MDC) with 43 layers. The average single wire
resolution is 135 ¦Ìm, and the momentum resolution for 1 GeV/c
charged particles in a 1 T magnetic field is 0.5\%; 2) an
electromagnetic calorimeter (EMC) made of 6240 CsI (Tl) crystals
arranged in a cylindrical shape (barrel) plus two endcaps. For 1.0
GeV photons, the energy resolution is 2.5\% in the barrel and 5\% in
the endcaps, and the position resolution is 6 mm in the barrel and 9
mm in the endcaps; 3) a Time-Of-Flight system (TOF) for particle
identification composed of a barrel part made of two layers with 88
pieces of 5 cm thick, 2.4 m long plastic scintillators in each
layer, and two endcaps with 96 fan-shaped, 5 cm thick, plastic
scintillators in each endcap. The time resolution is 80 ps in the
barrel, and 110 ps in the endcaps, corresponding to better than a 2
sigma $K/\pi$ separation for momenta below about 1 GeV/c; 4) a muon
chamber system (MUC) made of 1000 m$^2$ of Resistive Plate Chambers
(RPC) arranged in 9 layers in the barrel and 8 layers in the endcaps
and incorporated in the return iron of the superconducting magnet.
The position resolution is about 2 cm.

The estimation of physics backgrounds are performed through Monte
Carlo (MC) simulations. The GEANT4-based simulation software
BOOST~\cite{boost} includes the geometric and material description
of the BESIII detectors, detector response and digitization
models, as well as the tracking of the detector running conditions
and performance. The production of the $\jpsi$ resonance is
simulated by the MC event generator KKMC~\cite{kkmc}, while the
decays are generated by EvtGen~\cite{evtgen} for known decay modes
with branching fractions being set to the PDG~\cite{PDG} world
average values, and by Lundcharm~\cite{lundcharm} for the remaining
unknown decays. The analysis is performed in the framework of the
BESIII Offline Software System (BOSS)~\cite{boss} which takes care
of the detector calibration, event reconstruction and data storage.

\section {Event Selection}

The $\etap$ is identified by its decay into $\eta \pp$ with $\eta\to
\gamma \gamma$ in $\jpsi$ radiative decays, and candidate events
with the topology $\gamma \gamma \gamma \pip \pim $ are selected
using the following criteria.  Charged tracks in BESIII are
reconstructed from MDC hits. To optimize the momentum measurement,
we select tracks in the polar angle range $|\cos\theta|<0.93$ and
require that they pass within $\pm$10 cm of the interaction point in
the beam direction and within $\pm$1 cm in the plane perpendicular
to the beam. Electromagnetic showers are reconstructed by clustering
EMC crystal energies.  Efficiency and energy resolution are improved
by including energy deposits in nearby TOF counters. Showers
identified as photon candidates must satisfy fiducial and
shower-quality requirements.  The minimum energy is 25 MeV for
barrel showers ($|\cos\theta|<0.8$) and 50 MeV for endcap showers
($0.86<|\cos\theta|<0.92$). Photons in the region between the barrel
and endcaps are not well measured and are not used. To exclude
showers from charged particles, a photon must be separated by at
least 20$^{\circ}$ from any charged track. EMC cluster timing
requirements suppress electronic noise and energy deposits unrelated
to the event.

The TOF (both Endcap and Barrel) and $dE/dx$ measurements for each
charged track are used to calculate $\chi_{PID}^2(i)$ values and the
corresponding confidence levels $Prob_{PID}(i)$ for the hypotheses
that a track is a pion, kaon, or proton, where $i~(i=\pi/K/p)$ is
the particle type. For pion candidates, we require
$Prob_{PID}(\pi)>Prob_{PID}(K)$ and $Prob_{PID}(\pi)>0.001$.

Candidate events must have two charged tracks with zero net charge,
and the number of photons should be greater than two. At least one
charged track must be identified as a pion. We do four-constraint (4C)
kinematic fits imposing energy and momentum conservation under the
$\jpsi \to \gamma \gamma \gamma \pi^+ \pi^-$ hypothesis looping over
all photon candidates, and select the combination with the minimum
$\chi^2(\gamma \gamma \gamma \pip \pi^-)$.  The minimum
$\chi^2(\gamma \gamma \gamma \pip \pi^-)$ should be less than 200, and the efficiency of this
requirement is around 99\%.
The $\eta$ candidates are
selected from the combination with the two photons' invariant mass
closest to $\eta$ nominal mass. With the above event selections, a
very clear $\eta$ signal is observed. In the analysis below, we define
the $\eta$ signal region as $0.518<m_{\gamma \gamma}<0.578$~GeV/$c^2$,
and the $\eta$ mass sidebands region as $0.443<m_{\gamma
\gamma}<0.473$~GeV/$c^2$ or $0.623<m_{\gamma \gamma
}<0.653$~GeV/$c^2$.


The backgrounds in the selected event sample from a number of
potential background channels listed in the PDG~\cite{PDG} are studied
with MC simulations.  The background level is very low in the $\etap$
mass region. The main backgrounds are from $\jpsi \to \gamma \etap \to
\gamma \gamma \rho^0 \to \gamma \gamma \pp$ and $\jpsi \to \gamma
\etap \to \gamma \gamma \omega \to \gamma \gamma \pp \pi^0$, which can
be described by the normalized $\eta$ mass sidebands events. The other
backgrounds with $\eta$ candidates are from $\jpsi \to \gamma
f_1(1285)/\eta(1405/1475)/f_1(1510)\to \gamma \eta \pp$.  None of
these backgrounds give peaking backgrounds in the $\etap$ mass region.
The total background contamination is estimated to be only 0.57\%
within the $\etap$ mass region ($\sim$3$\sigma$). An inclusive MC event
sample is also used to investigate other possible surviving background
events, but no other possible background from the inclusive MC is found.

\section{\boldmath Number of $J/\psi$ events}

The number of $J/\psi$ events, $N_{J/\psi}$, used in this analysis
is determined from the number of inclusive events. Charged tracks
are selected requiring their points of closest approach to the beam
line be within 15~cm of the interaction point along the beam line
and within 1~cm in the plane perpendicular to the beam line, their
angles with respect to the beam line, $\theta$, must satisfy $|\cos
\theta| < 0.93$, and their momenta must be less than 2.0 GeV/$c$.
Clusters in the EMC must have at least 25 (50) MeV of energy in the
barrel (endcap) EMC, and have $|\cos \theta| < 0.93$.

Event selection requires at least two charged tracks and visible
energy, $E_{vis}$, greater than 1.0 GeV. Here $E_{vis}$ is defined
as the sum of charged particle energies computed from the track
momenta assuming pion masses, plus the neutral shower energies
measured in the EMC. To reduce backgrounds from Bhabha and dimuons,
events with only two charged tracks must have the momenta of the
charged tracks less than 1.5 GeV/$c$ and their energy deposit in the
EMC less than 1.0 GeV. Backgrounds from Bhabha and dimuon events
surviving the selection criteria are small. The continuum
contribution ($e^+ e^- \to$ anything) and the surviving backgrounds
are removed by subtracting the number of events selected with the
above criteria from a continuum sample taken at a center of mass
energy of 3.08 GeV and normalized by relative luminosity and the
cross section assuming a $1/s$ dependence.

The number of $\jpsi$ inclusive events is also determined from the
distribution of $\bar{z}$, which is the average of the $z$ distances
from the interaction point along the beam of the point of closest
approach of tracks to the beam line. Here the number of $\jpsi$
inclusive events is taken to be the number of events in a signal
region ($-4 < \bar{z} < 4$ cm) minus the number of events in
sideband regions ($ 6 < |\bar{z}| < 10$ cm) of $\bar{z}$.

\begin{figure}[htb]
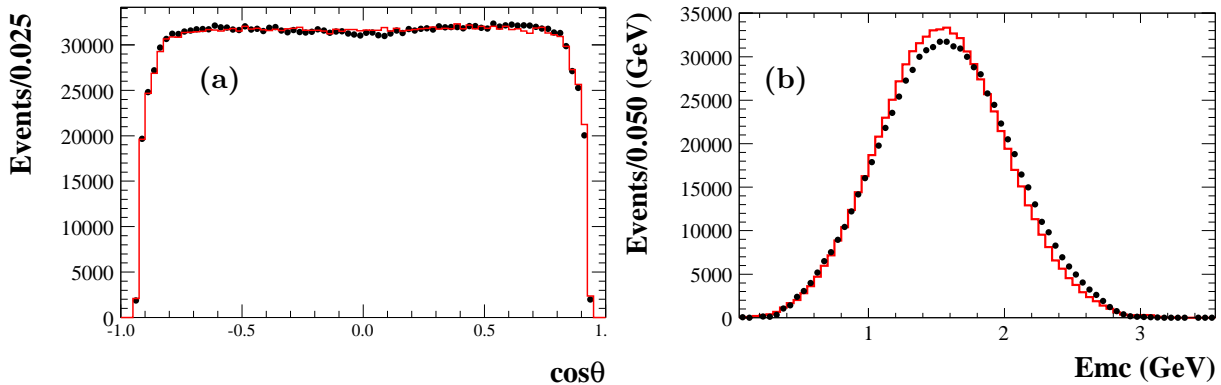

\centering
\includegraphics[width=5.0cm,angle=-90]{fig1a.epsi}
\includegraphics[width=5.0cm, angle=-90]{fig1b.epsi}
 \put(-385,-31){\bf (a)}
 \put(-175,-31){ \bf (b)}
\caption{(a) The $\cos \theta$ distribution of charged tracks for
events satisfying selection criteria. (b) The distributions of the
total energy in the EMC for events satisfying selection criteria.
Dots are data, and the histogram is $J/\psi \to inclusive$ simulated
events. } \label{costheta}
\end{figure}

The efficiency is determined from data using $J/\psi$ events from
$\psi' \to \pi^+ \pi^- J/\psi$ decays~\cite{Bai:1995ik} in the
BESIII 106 M $\psi'$ sample~\cite{npsip}. MC simulation is used to
determine a small correction (1.0108) to this efficiency arising
from the two extra tracks and the motion of the $J/\psi$ in the
$\psi'$ events. This procedure is less sensitive to differences
between data and MC simulation than using only MC to determine the
efficiency. The agreement between data and MC simulation is shown
for the $\cos \theta$ distribution of charged tracks in
Fig.~\ref{costheta}(a) and the total energy deposit in the EMC,
$E_{EMC}$, in Fig.~\ref{costheta}(b). The discrepancy between data
and MC simulations in Fig.~\ref{costheta}(b) is due to the
imperfect MC generator and imperfect detector simulation.
The systematic error due to the $E_{vis}$ requirement is negligible.
\vspace{0.3cm}

The result is $N_{J/\psi} = (225.2 \pm 2.8)\times 10^6$, where the
error is systematic and is determined mostly by the track efficiency
difference between data and MC (0.41\%), the variation with the
minimum charged track multiplicity requirement (0.78\%), the
difference when the noise levels in the two samples of $\jpsi$ and
$\psi'$ events are modified (0.49\%), the error associated with
fitting the distribution of mass recoiling from the $\pi^+\pi^-$ to
determine the number of $\psi' \to \pi^+\pi^- J/\psi$ events
(0.45\%), the error due to the continuum subtraction (0.18\%), the
difference between the continuum subtraction and the sideband
subtraction methods for determining the number of events (0.18\%),
and the difference for changing the generator (0.49\%). The
statistical error is negligible.  A second analysis determines
$N_{J/\psi}$ from $J/\psi \to l^+ l^-$ events, where $l$ is a $\mu$
or $e$, and obtains consistent results.

\section {Branching fraction measurement}

\begin{figure}[htbp]
\includegraphics[width=6cm, angle=-90]{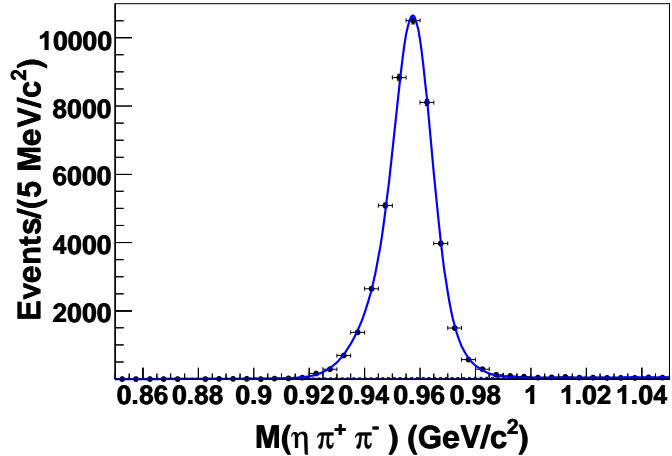}
\caption{ The
$\eta \pp$ invariant mass distribution of the final candidate
events. The dots with error bars represent data, and the solid curve
is the result of the fit described in the text. The dashed curve is
the background polynomial. } \label{metap-fit}
\end{figure}

Figure~\ref{metap-fit} shows the invariant mass distribution of
$\eta \pp$ candidate events.  This distribution is fitted with a
double-Gaussian function for the $\etap$ signal and a linear
function for the background shape. The fit yields $43826\pm 211$
events. The $\jpsi \to \gamma \etap$ branching fraction is
calculated using
$$
\BR(\jpsi \to \gamma \etap)=\frac{N^{obs}}{N_{\jpsi}\times
\eff\times \BR(\etap \to \eta \pp) \times \BR (\eta \to \gamma
\gamma)},
$$
where $N^{obs}$ is the number of events observed, $N_{\jpsi}$ is the
number of $\jpsi$ events, and $\eff$ is the selection efficiency
obtained from MC simulation, which is 23.57\%. The branching
fraction is then determined to be $(4.84\pm 0.03)\times 10^{-3}$,
where the error is statistical only. We also check $\BR(\jpsi \to \gamma \etap)$
by using the number of events after the 6C kinematic fit requirement (the reconstructed momenta of
two gammas are constrained to the $\eta$ mass and the reconstructed
momenta of $\eta \pp$ is constrained to the $\etap$ mass), where the number
of signal events is obtained by subtracting all the simulated normalized backgrounds
with $\eta$ candidates and normalized $\eta$ mass sidebands events directly.
The difference for $\BR(\jpsi \to \gamma \etap)$ is only 0.3\%.

\section{Measurement of the Matrix Element}

The internal dynamics of the decay $\etap \to \eta \pp$ can be
described by two degrees of freedom since all the particles are spin
zero particles. The Dalitz plot distribution for the charged decay
channel $\etap \to \eta \pp$ is described by the following two
variables:
\begin{equation}
X=\frac{\sqrt{3}}{Q}(T_{\pi^+}-T_{\pi^-}), \hspace{2cm}
Y=\frac{m_{\eta}+2m_{\pi}}{m_{\pi}}\frac{T_{\eta}}{Q}-1,
\end{equation}
where $T_{\pi, \eta}$ denote the kinetic energies of mesons in the
$\etap$ rest frame and
$Q=T_{\eta}+T_{\pip}+T_{\pim}=m_{\etap}-m_{\eta}-2m_{\pi}$. The
squared absolute value of the decay amplitude is expanded around the
center of the corresponding Dalitz plot in order to obtain the
Dalitz slope parameters:
\begin{equation}\label{equa1}
M^2=A(1+aY+bY^2+cX+dX^2),
\end{equation}
where $a,b,c$ and $d$ are real parameters and A is a normalization
factor. This parametrization is called the general decomposition.
The parametrization in Eq.~(\ref{equa1}) has also been proposed with
an extra term, either $eXY$ or $fX^3+gY^3$.  For the charged channel
$\etap \to \eta \pp$, odd terms in $X$ are forbidden due to charge
conjugation symmetry, while for the neutral channel $\etap \to \eta
\pi^0 \pi^0$, $c=0$ from symmetry of the wave function. The Dalitz
plot parameters may not be necessarily the same for charged and
neutral decay channels. However, in the isospin limit they should be
the same.

A second parametrization is the linear one~\cite{PDG}:
\begin{equation}\label{equa2}
M^2=A(|1+\alpha Y|^2+cX+dX^2),
\end{equation}
where $\alpha$ is a complex parameter. Of particular interest is the
real component of the complex constant $\alpha$, which is a linear
function of the kinetic energy of the $\eta$. A non-zero value of
$\alpha$ may represent the contribution of a gluon component in the
wave function of the $\etap$ in the dynamics of its
decay~\cite{gams}. Comparison with the general parametrization gives
$a = 2\hbox{Re}(\alpha)$ and $b =
\hbox{Re}^2(\alpha)+\hbox{Im}^2(\alpha)$. Both parameterizations are
equivalent if $b > a^2/4$.

To improve the $\eta$ and $\etap$ mass
resolutions and reduce the migration of events to the nearby bins in the Dalitz plot, we use kinematic information after a 6C kinematic fit to
calculate the $X$ and $Y$ values.

\begin{figure}[htbp]
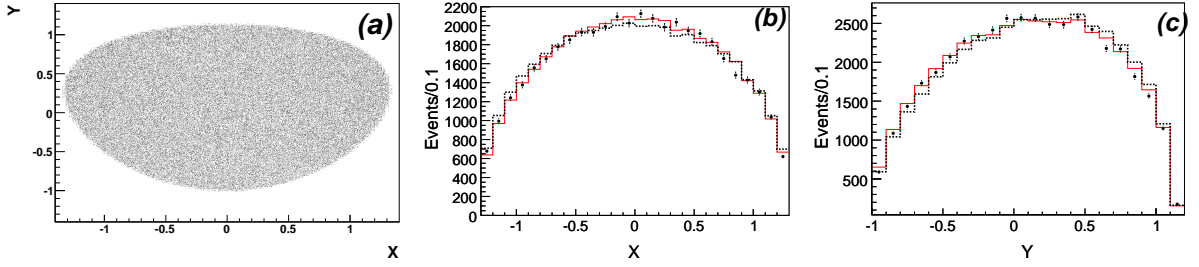

\includegraphics[width=3.4cm, angle=-90]{fig3a.epsi}\hspace{0.2cm}
\includegraphics[width=3.4cm, angle=-90]{fig3b.epsi}\hspace{0.2cm}
\includegraphics[width=3.4cm, angle=-90]{fig3c.epsi}
\caption{(a) The
experimental Dalitz diagram for the decay $\etap \to \eta \pp$ in
terms of the variables $X$ and $Y$ with the $\eta \pp$ mass in the
$\etap$ mass region. The corresponding projections on variables $X$
and $Y$ are shown in (b) and (c), respectively, where the dashed
histograms are from MC signal sample with $\etap \to \eta \pp$ events
produced with phase space and the solid histograms are the fitted
results described in the text.} \label{xy}
\end{figure}

Figure~\ref{xy} (a) shows the experimental form of the Dalitz diagram
for the decay $\etap \to \eta \pp$ in terms of the variables $X$ and
$Y$ with the $\eta \pp$ mass in the 0.93-0.98 GeV/$c^2$ mass region,
while the corresponding projections on variables $X$ and $Y$ are shown
in Figs.~\ref{xy} (b) and (c), respectively. In Figs.~\ref{xy} (b) and
(c), the dashed histograms are from MC signal sample with $\etap \to
\eta \pp$ events produced with phase space, while the solid histograms
are the fitted results described below.  The resolutions in the
variables $X$ and $Y$ over the entire 6C kinematical region are
$\sigma_X=0.03$ and $\sigma_Y = 0.025$, respectively, according to MC
simulation.

\begin{figure}[htbp]
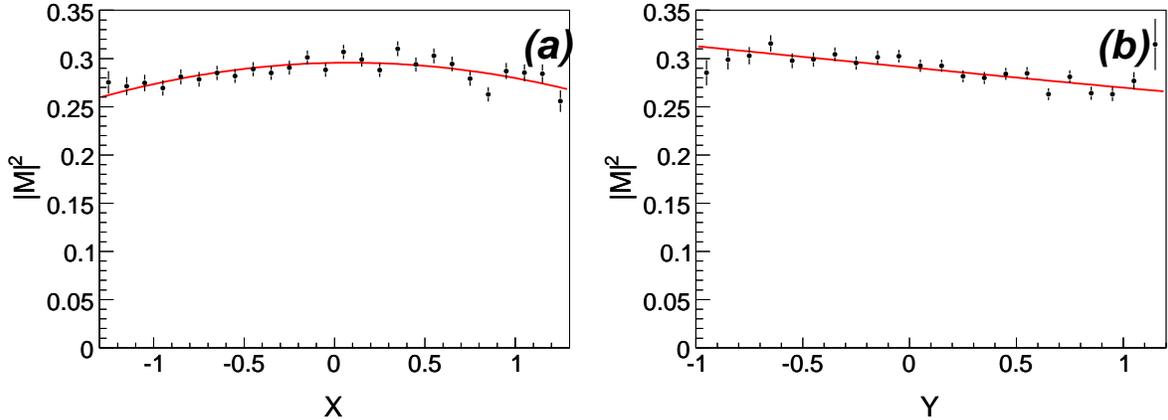

\includegraphics[width=5.5cm, angle=-90]{fig4a.epsi}\hspace{0.3cm}
\includegraphics[width=5.5cm, angle=-90]{fig4b.epsi}
\caption{Dependence
of the square of the $\etap$ decay matrix element on the Dalitz
variables $X$ and $Y$. The solid lines are the results of the fits
of the data described in the text. } \label{xy-1d}
\end{figure}

The dependence of the matrix element on each variable, $X$ and $Y$,
after integration over the other, and after dividing by phase space,
is shown in Fig.~\ref{xy-1d}. Fitting the data with
Eq.~(\ref{equa2}) gives the following values of the parameters:
$\hbox{Re}(\alpha)= -0.035 \pm 0.005$, $\hbox{Im}(\alpha)=0.00\pm
0.08$, $c=0.018\pm 0.008$ and $d=-0.059 \pm 0.012$, where the errors
are statistical only. Although the fitted results are consistent
with world average values~\cite{PDG}, possible correlations between
the $X$ and $Y$ are not considered. Also the fitted value of the
parameter $d$ is not consistent with zero so the matrix element can
not be well described by a linear function of $Y$ only. So we do the
fits to the Dalitz plot described below.

In the fitting procedure, the Dalitz
plot is subdivided into 26 $X$-bins and 22 $Y$-bins, i.e. 572 cells
in total. Dalitz plot parameters are obtained by
minimization of the function:
\begin{equation}
\chi^2(N, a, b, c, d)= \sum_{i}^{n_{bin}}\frac{(D_i - N
M_i)^2}{\sigma_{i}^2}
\end{equation}
Here the index $i$ enumerates cells in Dalitz plot (empty
cells outside the Dalitz plot boundaries are excluded), N is
normalization factor, $a, b, c$ and $d$ are the Dalitz plot
parameters. The $M_i$ and $D_i$ are the numbers of (weighted)
entries in the $i$-th bin of the two-dimensional histograms in the
Dalitz variables for MC and for the background-subtracted data,
respectively. The statistical error $\sigma$ includes background
subtraction and MC statistical errors. The MC histogram is obtained
as follows:
\begin{equation}
M_i=\sum_{j=1}^{N_{ev}}(1+aY_j+bY_j^2+cX_j+dX_j^2),
\end{equation}
for the general decomposition parametrization, where the index $j$ is over
the generated events and $X_j$ and $Y_j$ are the generated true values of
Dalitz variables. Similarly for the linear parametrization,
\begin{equation}
M_i=\sum_{j=1}^{N_{ev}}(|1+\alpha Y_j|^2+cX_j+dX_j^2).
\end{equation}
The fit procedure has been verified with MC by checking the input and
output values of the Dalitz plot parameters.

\begin{figure}[htbp]
\includegraphics[width=10cm, angle=-90]{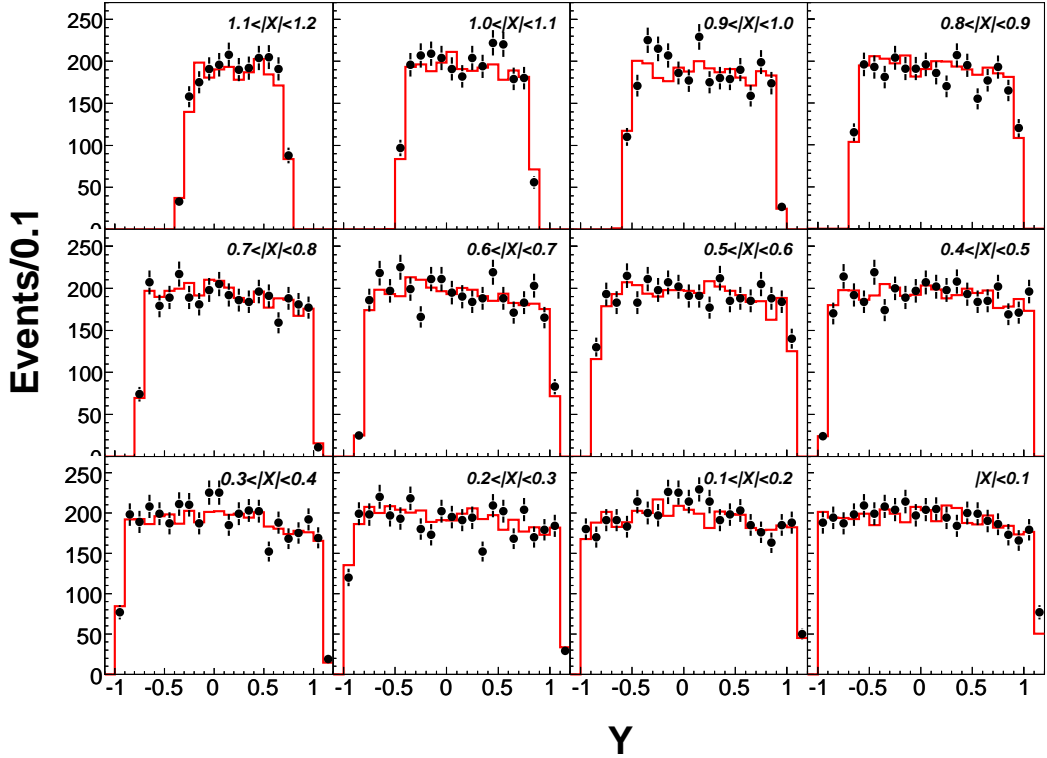}
\caption{Experimental
distributions of the variable $Y$ in various intervals of $X$ with the
fitting function (histogram) for the general decomposition
parametrization.} \label{y-fit}
\end{figure}

First we fit using the general decomposition
parametrization of the matrix element and obtain the
following values for the parameters of the matrix element and for
the correlation matrix ($\chi^2/NDF=504/476$, where $NDF$ is the
number of degrees of freedom.):
\begin{equation}
\begin{matrix}
a=-0.047\pm0.011 \\
b=-0.069\pm0.019 \\
c=+0.019\pm 0.011 \\
d=-0.073\pm0.012
\end{matrix}
\begin{pmatrix}
1.000 & -0.442  & -0.010 & -0.239\\
& 1.000 & 0.025 & 0.282 \\
& & 1.000 & 0.030 \\
& & & 1.000
\end{pmatrix}
\end{equation}
The errors are statistical only. This result is illustrated in
Fig.~\ref{y-fit}, where we show the comparison of data (dots with
error bars) and MC weighted with fitted coefficients (histogram) as a
function of $Y$, in different $X$-intervals for $\etap \to \eta
\pp$. Parameter $c$ is consistent with zero within 1.8$\sigma$. The
fitted results are almost the same with the value of parameter $c$
fixed at zero. The statistical significance of $c$ is estimated to be
$2.1$$\sigma$, from the difference of the $\chi^2$ value taking the
difference in the number of degrees of freedom ($\Delta \hbox{NDF}=1$)
in the fits into account.

The extra term $eXY$ or $fX^3+gY^3$ has also been added into the
general parametrization. The fitted value of parameter $e$ is
$0.000\pm0.018$, which is consistent with the conclusion from the VES
measurement~\cite{ves1}. The fitted results of parameters f and g
are $0.037\pm0.035$ and $-0.014\pm0.018$, respectively, and the
corresponding statistical significances are very small ($\sim
1$$\sigma$). All the other parameter values are almost the same.

We also perform a fit using the linear
parametrization of the matrix element and obtain
($\chi^2/NDF=521/476$):
\begin{equation}
\begin{matrix}
\hbox{Re}(\alpha)=-0.033\pm0.005 \\
\hbox{Im}(\alpha)=0.000\pm0.049 \\
c=+0.018\pm 0.009 \\
d=-0.059\pm0.012
\end{matrix}
\begin{pmatrix}
1.000 & -0.001  & 0.001 & -0.138\\
& 1.000 & 0.000 & 0.000 \\
& & 1.000 & 0.024 \\
& & & 1.000
\end{pmatrix}
\end{equation}
The errors are statistical only.  The parameter $c$ is consistent with
zero within 2.0$\sigma$, and the statistical significance  is
estimated to be $2.2$$\sigma$.

\section{Systematic Uncertainties}

The sources of the systematic errors for the branching fraction
measurement are summarized in Table~\ref{errbr}. The uncertainty is
negligible for pion identification since the identification of only
one of the pions is required. The uncertainty in the tracking
efficiency is 1\% per track and is additive. The uncertainty
associated with the kinematic fit is determined to be 0.2\% using
the control sample $\jpsi \to \pp \pi^0$. The uncertainty due to
photon detection is 1\% per photon. This is determined from studies
of photon detection efficiencies in well understood decays such as
$\jpsi \to \rho^0 \pi^0$ and study of photon conversion via $\EE \to
\gamma \gamma$. According to the MC simulation, the trigger
efficiency for signal events is almost 100\%, and the uncertainty is
neglected. The background uncertainties are evaluated by changing
the background fitting function from a first order polynomial to
second order and the fitting range, resulting in change of the
branching fraction by 0.3\%. The uncertainties of $\BR(\etap \to
\eta \pp)$ and $\BR(\eta \to \gamma \gamma)$ are 1.6\% and 0.5\%,
respectively~\cite{PDG}. The fitted results to the Dalitz plot
matrix element show correlations between the Dalitz plot parameters.
This should be properly taken into account when integrating the
amplitude over phase space to obtain the decay width. The maximum
difference in the efficiency is 2.5\% by using general decomposition
parametrization results or linear parametrization results. The
difference (2.5\%) is conservatively taken into the systematic
errors. Finally the uncertainty on the number of $\jpsi$ events is
2\%. Assuming that all of these systematic error sources are
independent, we add them in quadrature to obtain the total
systematic error shown in Table~\ref{errbr}.

\begin{table}[htbp]
\caption{Relative systematic errors (\%) for the branching fraction
measurement.} \label{errbr}
\begin{tabular}{c | c }
\hline Source & $\BR(\jpsi \to \gamma \etap)$
\\\hline
 Part ID &  ---  \\
 Tracking & 2.0  \\
 Kinematic fit & 0.2 \\
 Photon efficiency & 3.0 \\
 MC statistics & 0.3 \\
 Trigger efficiency &  --- \\
 Background shape & 0.3  \\
 Intermediate branching fractions & 1.7 \\
 Dalitz plot matrix element & 2.5  \\
 Number of $\jpsi$ events & 1.3 \\
 \hline
 Sum in quadrature & 4.9 \\
 \hline
\end{tabular}
\end{table}

The systematic errors in the measurement of the Dalitz plot matrix
element are summarized in Table~\ref{err_parameter}.  The uncertainty
from the backgrounds is negligible since the contamination is very
small ($\sim$0.57\%). The tracking efficiency correction functions for
$\pip$ and $\pim$ are obtained by using the control sample $\jpsi \to
\pp p \bar{p}$, where the transverse momentum region of pion has
covered the region of signal pion transverse momentum. The differences
on the fitted values of parameters $a$, $b$, $c$ and $d$ are 3.3\%,
3.3\%, 4.4\% and 1.2\% in the general parametrization, and 3.3\%,
4.8\%, 2.4\% for the parameters $\hbox{Re}(\alpha)$, c and d in the
linear parametrization by applying the tracking efficiency correction
functions for $\pip$ and $\pim$, respectively. The differences on the
fitted results of parameters due to changing the $\eta \pp$ mass
requirement are included in the systematic errors. The fitted results
are also compared using a 4C instead of the 6C kinematic fit, and the
corresponding differences are taken as the systematic errors due to
the kinematic fit uncertainty. Binning size was changed up to a factor
of two: $0.1<\Delta X, \Delta Y<0.2$. The biggest differences on the
fitted parameter values are taken as the systematic errors due to the
binning size uncertainty. To determine the systematic errors associated
with the event selection, especially for the selection of $\eta$
candidates, another set of event selection criteria are applied: (1)
The photon with the maximum energy is regarded as the radiative photon
($\gamma_{rad}$). (2) We do a 4C kinematic fit to the $\jpsi \to
\gamma_{rad} \gamma \gamma \pip \pi^-$ hypothesis looping over all the
other photon candidates and select the combination with the minimum
$\chi^2(\gamma_{rad} \gamma \gamma \pip \pi^-)$. (3) The $\gamma
\gamma$ mass is required to be within the $\eta$ mass region. After
applying the above event selection criteria, the difference on the
total number of signal events is only about 0.53\%. The fits to the
Dalitz plot parameters are done with these events, and the differences
are included into the systematic errors due to the event selection
method uncertainty. The PHOTOS package~\cite{photos} was used to
include final state radiation (FSR). By changing the ratio of FSR
events, the differences are taken into the systematic errors due
to the FSR simulation uncertainty. Assuming that all the sources are
independent and adding them in quadrature, one gets the total
systematic errors of parameters $a$, $b$, $c$ and $d$ in the generalized
representation are 4.9\%, 12\%, 12\% and 3.1\%, and the total
systematic errors of parameters $\hbox{Re}(\alpha)$, $c$ and $d$ are
7.2\%, 12\% and 6.5\%  in the linear representation, respectively.

\begin{table}[htbp]
\caption{Relative errors of the parameters of the matrix element for
the generalized and linear representations.} \label{err_parameter}
\begin{tabular}{c | c c c c | c c c c  }
\hline Source & \multicolumn{4}{c|}{generalized representation} &
\multicolumn{3}{c}{linear representation}\\
 & $a$ & $b$ & $c$ & $d$ & $\hbox{Re}(\alpha)$  & $c$ &
 $d$\\\hline
 Tracking efficiency & 3.3 & 3.3 & 4.4 & 1.2  & 3.3 & 4.8 & 2.4
 \\
 $m_{\eta \pp}$ mass cut & 0.9 & 4.8 & 3.3 &  1.7 & 1.4 & 2.1 & 3.2
 \\
 Kinematic fit & 2.8 & 4.9& 2.1& 0.7&  5.2& 7.0& 4.7
\\
 Binning size& 0.9 & 8.0 & 9.2 & 1.5& 2.6& 5.7& 1.7 \\
 Different selection method & 1.6 & 2.9 & 0.7 & 1.4 & 2.0 & 4.8 &
 1.1\\
 FSR simulation & 1.0 & 0.4 & 1.9 & 0.2& 0.9& 0.7& 0.4
 \\\hline
Sum in quadrature & 4.9 & 12 & 12 & 3.1 & 7.2 & 12 & 6.5 \\\hline
\end{tabular}
\end{table}

\section{summary}

Using the large $\jpsi$ sample ($(225.2\pm2.8)\times 10^6$ $\jpsi$
events) collected with BESIII, the branching fraction of $\jpsi \to
\gamma \etap$ is measured to be
$$
\BR(\jpsi \to \gamma \etap)=(4.84\pm0.03~({\rm stat})\pm0.24~({\rm
sys}))\times 10^{-3},
$$
which is consistent with the recent BESII value ($(5.55\pm 0.44)\times 10^{-3}$)~\cite{bes2-br} within 1.5$\sigma$, and the CLEO value
($(5.24\pm 0.17)\times 10^{-3}$)~\cite{cleo-br} within
1.4$\sigma$, which are used in obtaining the world average value $(5.28\pm 0.15)\times 10^{-3})$
by the PDG~\cite{PDG}.

The parameters of the matrix element for the decay process $\etap
\to \eta \pp$ have been determined for the generalized and linear
representations. They are:

$$
\begin{matrix}
a=-0.047\pm0.011\pm0.003 \\
b=-0.069\pm0.019\pm0.009 \\
c=+0.019\pm 0.011\pm0.003 \\
d=-0.073\pm0.012\pm0.003
\end{matrix}
$$
for the generalized parametrization, and
$$
\begin{matrix}
\hbox{Re}(\alpha)=-0.033\pm0.005 \pm 0.003 \\
\hbox{Im}(\alpha)=0.000\pm0.049 \pm 0.001 \\
c=+0.018\pm 0.009 \pm 0.003\\
d=-0.059\pm0.012 \pm 0.004
\end{matrix}
$$
for the linear parametrization, where the first errors are
statistical and the second systematic.

Table~\ref{tt} shows the experimental and theoretical values of the
parameters of the matrix element squared for $\etap \to \eta \pp$ in
the general parametrization (second, third and fourth columns) and in
the linear parametrization (sixth, seventh and eighth columns).
The theoretical values in Ref.~\cite{th1} are the latest calculations within
the framework of $U(3)$ chiral effective field theory in combination
with a relativistic coupled-channels approach.
We see: (1) The errors of our fitted parameter values are smaller
compared to previous published results.  (2) In the general
decomposition parametrization of the matrix element, the central
values of parameters $a$ and $b$ are consistent with the results
from GAMS-4$\pi$ Collaboration~\cite{gams}, where the neutral decay
$\etap \to \eta \pi^0 \pi^0$ events were analyzed; however the
central values of parameters $c$ and $d$ are consistent with the
results from VES Collaboration~\cite{ves1}. (3) The negative value of
the coefficient $b$ indicates that the two kinds of parametrization are
not equivalent. This conclusion is consistent with that from
GAMS-4$\pi$ Collaboration~\cite{gams}; however it is different from
the conclusion from the VES Collaboration~\cite{ves1}, where the fit
with linear parametrization yields an unsatisfactory
$\chi^2/NDF=170.5/114$ ratio. (4) The quadratic term in $X$ is
unambiguously different from zero. Similarly for the quadratic term
in $Y$.
The measured value of the $Y$-variable quadratic term (b) is not consistent
with the expected value of around zero in the Effective Chiral Lagrangian model
in which the lowest lying scalar meson
candidates $\sigma$ and $\kappa$ together with the $f_0(980)$ and $a_0(980)$ are combined
into a possible nonet~\cite{ch5}; however it can be accommodated in a
$U(3)$ Chiral Unitarized model by including  final state interactions~\cite{ch2}.
The dynamical nature of this term needs further clarification. (5) The
value of parameter $c$, which tests C parity violation in the strong
interaction,  is
consistent with zero within 2$\sigma$ in both
parametrizations. In the future, with
much more BESIII data, the hadronic decays of $\etap$ can 
be measured with higher precision, especially the Dalitz
decay parameters, allowing more stringent testing of the
predictions of ChPT~\cite{lihb}.

\begin{table}[htbp]
\caption{Experimental and theoretical values of the parameters of
the matrix element squared for $\etap \to \eta \pp$ in the general
parametrization (second, third and fourth columns) and in the linear
parametrization (sixth, seventh and eighth columns). } \label{tt}
{\footnotesize
\begin{tabular}{c c c c | c c c c  }
 \hline Par. & VES~\cite{ves1} & Theory~\cite{th1} &
This work & Par. & CLEO~\cite{cleo} & VES~\cite{ves2} & This work
\\\hline
a & $-0.127\pm0.018$ & $-0.116\pm0.011$ & $-0.047\pm0.012$  & $\hbox{Re}(\alpha)$ &$-0.021\pm0.025$ & $-0.072\pm0.014$ & $-0.033\pm0.006$ \\
b & $-0.106\pm0.032$ & $-0.042\pm0.034$ & $-0.069\pm0.021$ & $\hbox{Im}(\alpha)$ &  $0.000$ (fixed)  & $0.000\pm0.100$ & $0.000\pm0.050$\\
c & $+0.015\pm0.018$ & -- & $+0.019\pm0.012$ & c & $0.000$ (fixed)& $+0.020\pm0.019$ & $+0.018\pm0.010$ \\
d & $-0.082\pm0.019$ & $+0.010\pm0.019$ & $-0.073\pm0.013$ & d & $0.000$ (fixed) & $-0.066\pm0.034$ & $-0.059\pm0.013$ \\
 \hline
\end{tabular}}
\end{table}

\section{acknowledgments}

The BESIII collaboration thanks the staff of BEPCII and the computing center for their hard efforts. This work is supported in part by the Ministry of Science and Technology of China under Contract No. 2009CB825200; National Natural Science Foundation of China (NSFC) under Contracts Nos. 10625524, 10821063, 10825524, 10835001, 10935007; the Chinese Academy of Sciences (CAS) Large-Scale Scientific Facility Program; CAS under Contracts Nos. KJCX2-YW-N29, KJCX2-YW-N45; 100 Talents Program of CAS; Istituto Nazionale di Fisica Nucleare, Italy; Russian Foundation for Basic Research under Contracts Nos. 08-02-92221,  08-02-92200-NSFC-a; Siberian Branch of Russian Academy of Science, joint project No 32 with CAS; U. S. Department of Energy under Contracts Nos. DE-FG02-04ER41291, DE-FG02-91ER40682, DE-FG02-94ER40823; University of Groningen (RuG) and the Helmholtzzentrum fuer Schwerionenforschung GmbH (GSI), Darmstadt; WCU Program of National Research Foundation of Korea under Contract No. R32-2008-000-10155-0

\end{document}